\documentclass[showpacs,twocolumn,preprintnumbers,amsmath,amssymb,aps,prl,superscriptaddress]{revtex4}

\usepackage[sort&compress]{natbib}
\usepackage{graphicx}
\usepackage{dcolumn}
\usepackage{bm}

\newcommand{\lsmo}{La$_{0.8}$Sr$_{0.2}$MnO$_3$}
\newcommand{\pzt}{Pb(Zr$_{0.2}$Ti$_{0.8}$)O$_3$}
\newcommand{\sto}{SrTiO$_3$}
\newcommand{\oC}{$^\mathrm{o}$C}

\begin{document}

\title{Origin of the magnetoelectric coupling effect in Pb(Zr$_{0.2}$Ti$_{0.8}$)O$_3$/La$_{0.8}$Sr$_{0.2}$MnO$_3$ multiferroic heterostructures}

\author{C. A. F. Vaz}
\email[Corresponding author. Email: ]{carlos.vaz@cantab.net}%
\affiliation{Department of Applied Physics and CRISP, Yale
University, New Haven, Connecticut 06520}%

\author{J. Hoffman}
\affiliation{Department of Applied Physics and CRISP, Yale
University, New Haven, Connecticut 06520}%

\author{Y. Segal}
\affiliation{Department of Applied Physics and CRISP, Yale
University, New Haven, Connecticut 06520}%

\author{J. W. Reiner}
\affiliation{Department of Applied Physics and CRISP, Yale
University, New Haven, Connecticut 06520}%

\author{R. D. Grober}
\affiliation{Department of Applied Physics and CRISP, Yale
University, New Haven, Connecticut 06520}%

\author{Z. Zhang}
\affiliation{Advanced Photon Source, Argonne National Laboratory,
Argonne, Illinois 60439, USA}%

\author{C. H. Ahn}
\affiliation{Department of Applied Physics and CRISP, Yale
University, New Haven, Connecticut 06520}%

\author{F. J. Walker}
\affiliation{Department of Applied Physics and CRISP, Yale
University, New Haven, Connecticut 06520}%

\date{\today}

\begin{abstract}
The electronic valence state of Mn in
Pb(Zr$_{0.2}$Ti$_{0.8}$)O$_3$/La$_{0.8}$Sr$_{0.2}$MnO$_3$
multiferroic heterostructures is probed by near edge x-ray
absorption spectroscopy as a function of the ferroelectric
polarization. We observe a temperature independent shift in the
absorption edge of Mn associated with a change in valency induced by
charge carrier modulation in the La$_{0.8}$Sr$_{0.2}$MnO$_3$,
demonstrating the electronic origin of the magnetoelectric effect.
Spectroscopic, magnetic, and electric characterization shows that
the large magnetoelectric response originates from a modified
interfacial spin configuration, opening a new pathway to the
electronic control of spin in complex oxide materials.
\end{abstract}




\maketitle

Understanding how to couple the electric and magnetic order
parameters in the solid state is a long-standing scientific challenge
that is intimately linked to the spatial and temporal symmetries
associated with charge and spin. Coupling of the order parameters is
observed in many different materials, but the effect is generally
weak in magnitude, even in materials that are both ferroelectric and
ferromagnetic ({\it multiferroic}) \cite{Schmid94,Hill00,RS07}.
Increasing the magnitude of the coupling is a fundamental problem in
condensed matter physics with important implications for
applications. For example, strong magnetoelectric coupling allows
for the ultra-sensitive measurement of weak magnetic fields, and at
smaller length scales, enables spin-based technologies by allowing
the control of the spin state at the atomic scale via electric
fields.

In single phase multiferroics, the magnetic and ferroelectric orders
often occur largely independent of each other, and as a result the
magnetoelectric coupling tends to be small \cite{Hill00,Khomskii09}.
In order to overcome this intrinsic limitation in the coupling
between the order parameters, artificially structured materials with
enhanced magnetoelectric couplings have been engineered, where a
break in time reversal and spatial symmetry occurs naturally at the
interface between the different phases \cite{RS07,EMS06,RSS08}.
Moreover, the coupling mechanism can be tailored to benefit from
several phenomena, including elastic \cite{Fiebig05,TDB+07},
magnetic exchange bias \cite{BHC+05,LSM+06,CMH+08}, and charge-based
\cite{MHV+09} couplings. In charge-based multiferroic composites,
the sensitivity of the electronic and spin state of strongly
correlated oxides to charge provides enhanced coupling between
magnetic and ferroelectric order parameters \cite{MHV+09}; it often
relies on charge doping of a ``colossal'' magnetoresistive (CMR)
manganite to modulate between high and low spin states, which
compete for the ground state of the system. However, the microscopic
origin of this effect is still not fully understood. In particular,
the nature of the effect and how the interplay between charge, spin,
and valency combines to yield the large magnetoelectric response in
this system remain to be addressed. In this Letter, we explore the
sensitivity of x-ray absorption near edge spectroscopy (XANES) to
the atomic electronic state to demonstrate the microscopic origin of
the magnetoelectric effect found in \pzt/\lsmo\ (PZT/LSMO)
heterostructures. XANES is particularly sensitive to changes in the
atomic valence state, and therefore especially suited to probing the
atomic structure modifications in LSMO induced by change in the
charge carrier concentration; another key advantage is the ability
to probe buried layers. Direct quantification of the charge-driven
magnetic changes based on the spectroscopic, electric and magnetic
measurements show that both the spin state and spin configuration of
LSMO are modulated, whereby the interfacial spin coupling changes
from ferromagnetic to antiferromagnetic, giving rise to the large
magnitude of the magnetoelectric effect found in this system. A
similar interfacial magnetic reconstruction effect is also predicted
from first principles calculations in similar type of
heterostructures \cite{BT09}.

\begin{figure*}[t!]
\begin{centering}
\includegraphics*[width=17cm]{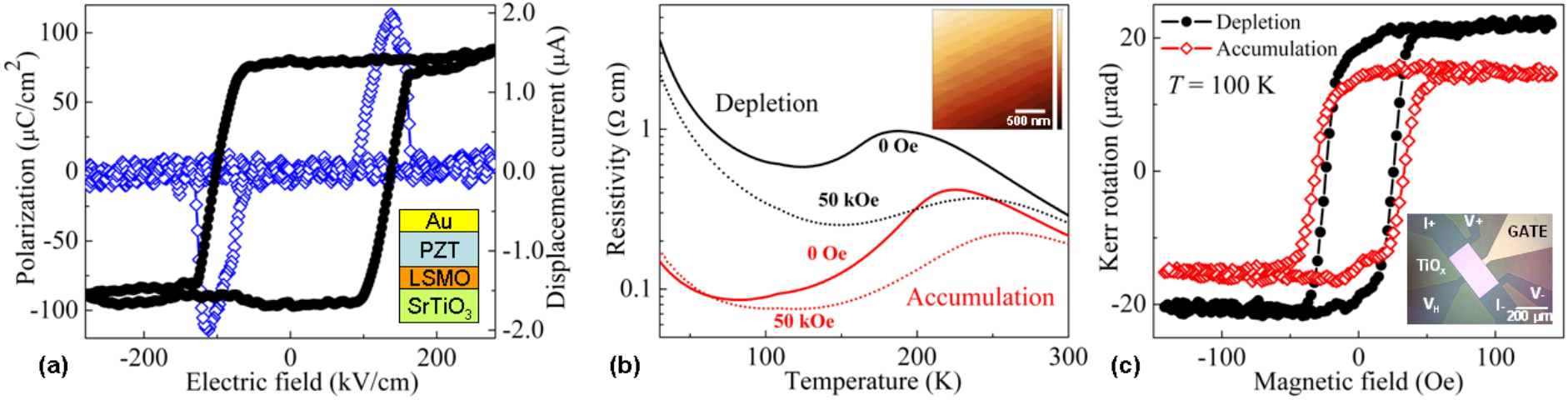}\\
\caption{Electric, transport and magnetic behavior of the PZT/12
u.c.\ LSMO structure, showing: (a) room temperature
polarization-electric field loop (full circles) and displacement
current (empty diamonds) of PZT. Inset: cross-sectional schematic of
the sample structure. (b) Resistivity versus temperature curves for
both the accumulation (red) and depletion (black) states at zero and
at 50 kOe applied magnetic field. Inset: atomic force microscopy
image of a 20 nm LSMO film (vertical scale range is 12 nm). (c)
Magnetic hysteresis curves of LSMO along the in-plane $\langle 100
\rangle$ direction for the two polarization states of PZT, taken at
100 K (averages of $\sim$300 individual traces). Inset: optical
image of the device before Au metallization; I+,I-,V+,V- denote the
current and voltage contacts for the resistivity measurements.}
\label{fig:M28}
\end{centering}
\end{figure*}

The samples in this study consist of  250 nm \pzt/ $t$
\lsmo/\sto(001) epitaxial heterostructures with $t = 11, 12$ unit
cells (u.c.), grown on TiO$_2$-terminated \sto\ substrates. LSMO is
a CMR oxide material characterized by a rich electronic phase
diagram, with properties that depend strongly on doping; at the
doping level chosen ($x=0.2$), the bulk LSMO system lies near the
boundary between insulating and metallic ferromagnetic ground states
\cite{IFT98,CDK+03a}. Two films were grown simultaneously for each
thickness, one on an unpatterned \sto\ substrate and another on a
TiO$_x$-masked substrate, exposing two identical Hall-bar device
structures defined by optical lithography. The active area is $160
\times 320$ $\mu$m$^2$ [Fig.~\ref{fig:M28}(c), inset]. The LSMO
films are grown by molecular beam epitaxy in an ultrahigh vacuum
deposition system with a base pressure of $1\times 10^{-10}$ mbar.
The elemental materials are evaporated from effusion cells under a
O$_2$ partial pressure of $1 \times 10^{-7}$ mbar, with the
substrates held at 720\oC. The evaporation rates are determined for
each material from a calibrated thickness monitor (2\%\ error),
while film thickness is monitored in real time using reflection high
energy electron diffraction (RHEED) intensity oscillations. The
observation of the latter indicates that film growth occurs in a
layer-by-layer mode; atomic force microscopy of LSMO films shows the
presence of unit cell high steps separating large, flat terraces
[see Fig.~\ref{fig:M28}(b), inset]. After the LSMO deposition, the
samples are cooled to room temperature in $1 \times 10^{-7}$ mbar of
O$_2$ and then transferred to an off-axis RF sputtering system for
PZT deposition using the conditions described in
Ref.~\onlinecite{HPLA03}. The LSMO films studied here have the
minimum thickness for which a peak in resistivity is observed
\cite{HPA05}; we focus here on the results obtained for 12 u.c.\
structures, which are representative of all measurements. The
direction of the PZT polarization is used to modulate
electrostatically the hole carrier density in the LSMO (hole charge
depletion occurs when the PZT polarization points down into the LSMO
film, and hole accumulation in the opposite case).

The XANES measurements were carried out at Beamline 33-ID of the
Advanced Photon Source, Argonne National Laboratory (Illinois). A
monochromatic x-ray beam $300 \times 120$ $\mu$m$^2$ in size
(defined using slits) was set perpendicular to the sample and
aligned onto the gate electrode of the device within 10 $\mu$m using
the measured x-ray induced photocurrent. The sample was mounted in
an evacuated variable temperature Displex cryostat equipped with a
Be dome. In this experiment, we probe the absorption K edge of Mn,
corresponding to the allowed electric dipole transition from the 1s
core level to unoccupied 4p states \cite{CSG+97,SGPB97,BBA+01}. The
light absorption, normalized to the incident photon flux, was
measured in the fluorescence mode using the Mn K$_\alpha$ line at
5899 eV using a Si drift x-ray detector spectrometer. Several scans
were performed on multiple device structures, including scan cycles
where the PZT was switched consecutively at each energy value (gate
voltage set to $\pm 5$ V). Scans were also carried out where the
x-ray absorption was measured at a fixed energy as a function of the
gate voltage. A reference x-ray absorption spectrum of metallic Mn
was taken for energy calibration.

The results of the electrical, transport and magnetic property
measurements of the PZT/LSMO structure are shown in
Fig.~\ref{fig:M28}. The electrical switching properties of the PZT
gate dielectric at room temperature, Fig.~\ref{fig:M28}(a), show
abrupt switching of the ferroelectric polarization. From integration
of the displacement current, we obtain a saturation polarization of
$P_s = 85$ $\mu$C/cm$^2$. Transport characteristics are shown in
Fig.~\ref{fig:M28}(b) for both states of the PZT polarization at a
magnetic field of 0 and 50 kOe, applied out of plane. The peak in
resistivity marks the transition between metallic and insulating
states, and is found to change by $\sim$40 K, from 188 K in the
depletion state to 226 K in the accumulation state. The presence of
a magnetic field leads to an increase in the resistivity peak
temperature, which is responsible for the CMR effect in the doped
manganites \cite{IFT98,Nagaev01}. The magnetic behavior was probed
with magnetooptic Kerr effect magnetometry in the longitudinal
geometry using an $s$-polarized HeNe laser beam ($\lambda = 633$ nm)
and a photoelastic modulator as part of an ellipsometer unit.
Magnetization versus temperature measurements (carried out as
described in Ref.~\onlinecite{MHV+09}) show a change in critical
temperature of 20 K, from 180 K in the depletion state to 200 K in
the accumulation state, showing that the peak in resistivity,
although correlated to the onset of magnetic order, does not
coincide exactly with the magnetic critical temperature
\cite{LBG+97a}. These results show that both the transport and
magnetic properties can be modulated electrostatically via the
ferroelectric field effect. For magnetic hysteresis loops
measurements, the magnetic field was varied linearly at a rate of 1
Hz, and a gate voltage of $\pm 10$ V was used to switch the PZT
polarization. The results are plotted in Fig.~\ref{fig:M28}(c) for
the two polarization states of PZT (100 K). The data show that the
accumulation state has a smaller saturation magnetization than the
depletion state, in agreement with previous results \cite{MHV+09}.
To obtain a quantitative estimate of the change in magnetization, we
carried out SQUID magnetometry on the unpatterned PZT/LSMO sample,
giving a saturation magnetization of 515 emu/cm$^3$, or
$m_\mathrm{dep} = 3.30$ $\mu_\mathrm{B}$/Mn, close to the bulk
value. Piezoelectric force microscopy of the unpatterned film shows
that the electric polarization state of the PZT after growth points
down; hence this value of the magnetic moment corresponds to the
LSMO in the depletion state. From the ratio
$m_\mathrm{dep}/m_\mathrm{acc} = 1.30(3)$ obtained from MOKE data,
we estimate the magnetic moment for the accumulation state as
$m_\mathrm{acc} = 2.54(6)$ $\mu_\mathrm{B}$/Mn, giving a change in
magnetic moment of $\Delta m = m_\mathrm{dep} - m_\mathrm{acc}=
0.76(6)$ $\mu_\mathrm{B}$/Mn. From these data, we obtain a
magnetoelectric coupling coefficient of $\Delta M/\Delta E =
6.2\times 10^{-3}$ $\mathrm{Oe\, cm\, V^{-1}}$ at 100~K.

\begin{figure}[t!]
\begin{centering}
\includegraphics*[width=8.5cm]{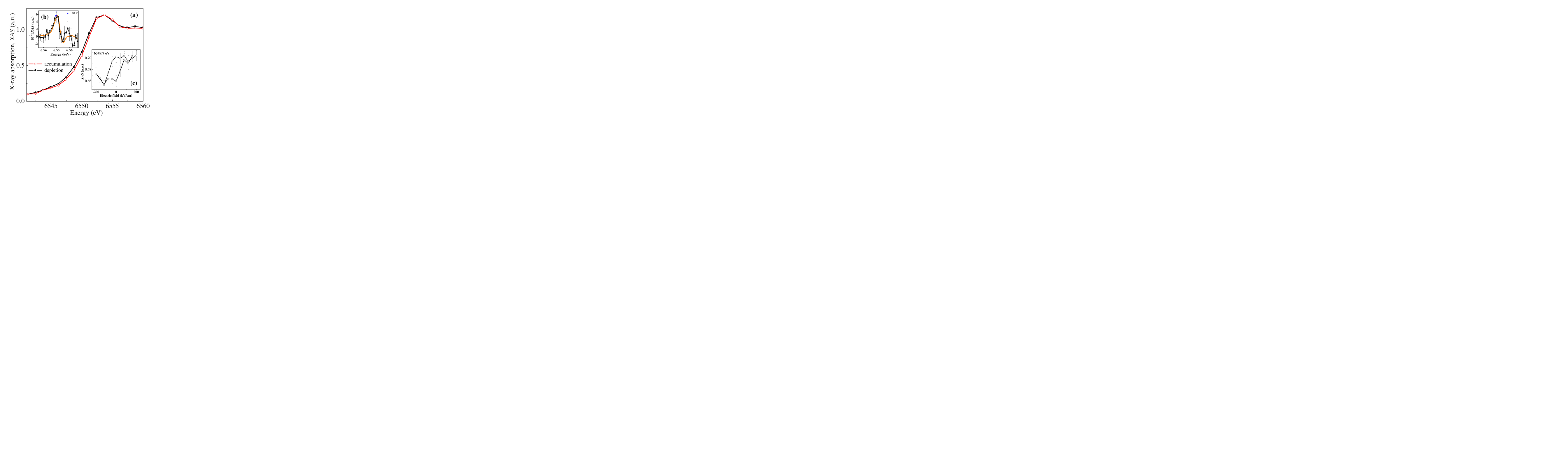}
\caption{(a) Room temperature XANES results for the two polarization
states of the PZT. (b) Difference in x-ray absorption for the two
PZT polarization states; the full line models this difference
assuming a rigid shift in the Mn absorption edge, taking for
reference the average between the accumulation and depletion states.
(c) Variation of the x-ray light absorption as a function of the
applied gate voltage at a fixed energy, $E=6549.7$ eV. The error
bars reflect counting statistics.} \label{fig:XASM28R}
\end{centering}
\end{figure}

The key experimental results of this work are the XANES data plotted
in Fig.~\ref{fig:XASM28R}, showing the room temperature x-ray
absorption response for the two states of the PZT polarization. The
main finding is the observation of an energy shift in the Mn
absorption edge by +0.3 eV upon switching the PZT polarization from
the depletion to the accumulation state. The position of the
absorption edge is very sensitive to the cationic valency in a wide
array of compounds \cite{KKP85}. In the case of the manganites, the
energy edge position is found to vary with the doping level $x$ by
about 3.5 eV in LaMn$_{1-x}$Co$_x$O$_3$ \cite{SKK+06}, 3-4.2 eV in
La$_{1-x}$Ca$_x$MnO$_3$ \cite{SGPB97,CSG+97,BBA+01}, 2.5-3 eV in
La$_{1-x}$Sr$_x$MnO$_3$ \cite{SBM03,BPK+05}, along with other more
subtle changes in peak amplitude and edge shape. The edge energy
position increases from the metallic state (Mn$^0$) to higher formal
valence states, and the observation of an energy shift between the
depletion and accumulation states shows directly a change in the
average Mn valency induced by the electrostatic hole-carrier
modulation. This change in the valency of the Mn is responsible for
the changes in the magnetic state of LSMO [Fig.~\ref{fig:M28}(c)]
and lies at the origin of the large magnetoelectric effect
\cite{MHV+09}. Importantly, in these measurements, no change in the
crystal lattice occurs. Measurements at low temperature (20 K)
display a similar change in the Mn valency, showing that the charge
carrier modulation is robust and occurs independent of temperature
[Fig.~\ref{fig:XASM28R}(b)]. The difference spectrum between
depletion and accumulation states is shown in
Fig.~\ref{fig:XASM28R}(b), which is reproduced well at the
absorption edge by a single parameter model corresponding to a rigid
shift in energy of 0.3 eV between the averaged absorption spectra. A
more direct visualization of the valence modulation can be seen in
Fig.~\ref{fig:XASM28R}(c), which shows the x-ray absorption as a
function of the applied gate voltage at fixed photon energy, showing
that the x-ray absorption can be modulated between a high and a low
value as the PZT polarization is switched. This result mimics the
{\it P-E} hysteresis curve [Fig.~\ref{fig:M28}(b)] and demonstrates
that the change in Mn valency tracks the switching of the PZT
polarization.

We may estimate the change in the Mn valency from XANES spectra for
the chemically doped LSMO compounds \cite{SBM03,BPK+05}. From the
results by Shibata et al.~\cite{SBM03} showing a linear variation in
the energy shift with the formal average valency of Mn, $\Delta E =
3.0\, x$, we estimate an averaged change in Mn valency of $\Delta x
= 0.1$/Mn between the accumulation and depletion states. The
relative change in the surface charge polarization of PZT, of $2P_s
= 1.6$ $e/\mathrm{u.c.}^2$, corresponds to a change in charge of
$\Delta n =0.13$/Mn. The agreement is striking and shows that most
surface charge is screened by charge carriers from the LSMO. This
change in valency is expected to take place mostly at the PZT/LSMO
interface, within the screening length of LSMO of about 1 u.c.\
\cite{HPA05}.

\begin{figure}[t!]
\begin{centering}
\includegraphics*[width=8.5cm]{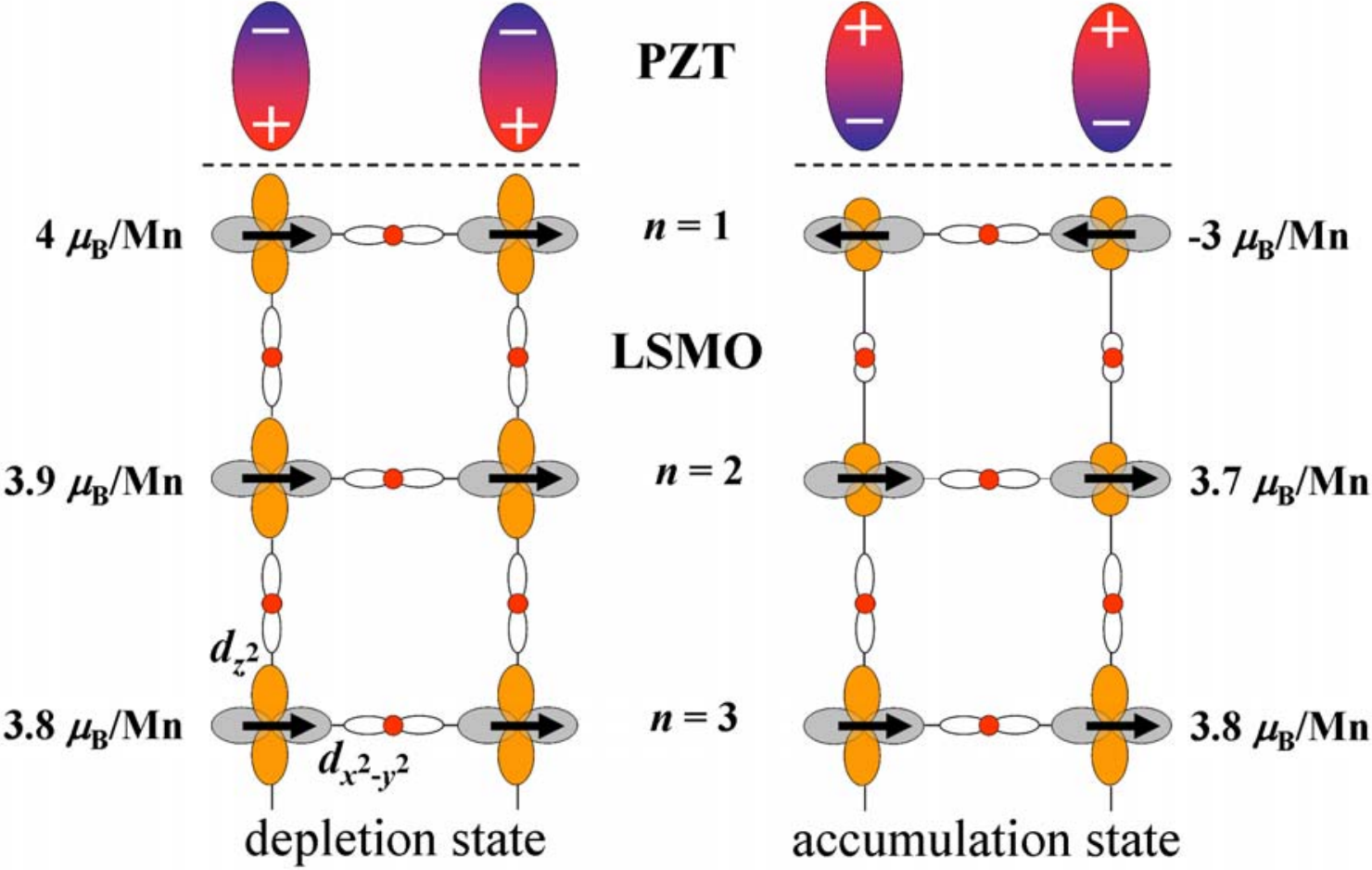}
\caption{Schematic model of the spin configurations in LSMO at the
PZT interface for the depletion and accumulation states, showing the
changes in the Mn and O orbital states and the expected changes in
the magnetic moment per layer. The arrows indicate the spin
orientation in the Mn cations and $n$ denotes the unit cell number
below the PZT. The Mn $d$ orbitals are drawn in orange and grey, and
the lobes of the $p$ orbitals are shown around the oxygen atoms
(red).} \label{fig:model}
\end{centering}
\end{figure}

One can now correlate the observed changes in the magnetization with
the valence modulation. Since going from the high spin state
Mn$^{3+}$ ($S=2$) to the low spin state Mn$^{4+}$ ($S=3/2$) changes
the magnetic moment by 1 $\mu_\mathrm{B}$, the measured change in Mn
valency gives a change in moment of 0.1 $\mu_\mathrm{B}$/Mn between
the depletion and accumulation states. Such a change is much smaller
than the measured change in magnetic moment of $\Delta m = 0.76(6)$
$\mu_\mathrm{B}$/Mn. Hence, the change in magnetic moment,
equivalent to about two u.c., must be explained by a mechanism other
than the change in the spin state of Mn induced by the valency
modulation. Instead, the change in moment must originate from a
modification in the spin exchange coupling at the interface, for
example, from ferromagnetic in the depletion state to
antiferromagnetic in the accumulation state, whereby the spins
couple ferromagnetically within this interfacial layer and
antiferromagnetically with the spins of the (001) adjacent layers.
Such spin arrangement has been recently calculated to be favored for
the accumulation state of La$_{0.5}$Ba$_{0.5}$MnO$_3$/BaTiO$_3$
multiferroic heterostructures \cite{BT09}. Our results provide an
experimental signature of the magnetic interface reconstruction
occurring at the PZT/LMSO interface. A model depicting the magnetic
reordering is shown in Fig.~\ref{fig:model}. In this picture, the
interface layer in the accumulation state consists of strongly
depopulated antibonding $e_g$ $3z^2 -r^2$ states, weakening the
double-exchange interaction at these orbitals; an antiferromagnetic
coupling to the adjacent layers would then ensue if the $x^2-y^2$
orbitals are energetically favored (favoring the superexchange
interaction). This result is predicted by first principles
calculations for LSMO under tensile strain, where a ferromagnetic
coupling is favored at low doping levels and antiferromagnetic
alignments are favored at higher doping \cite{FST00}. This change in
spin configuration results in a change in the average moment of
$(4+3+0.2)/12 = 0.6$ $\mu_\mathrm{B}$/Mn between depletion and
accumulation states (see Fig.~\ref{fig:model}), which is close to
the observed value. This result also agrees with the trend observed
in the bulk phase diagram, where an antiferromagnetic coupling of
alternating (001) planes is favored for hole doping between 0.5 and
0.65 at low temperatures \cite{IFT98,CDK+03a}. This mechanism gives
rise to a much more dramatic change in the average magnetic moment
and explains the very large magnetoelectric coupling
\cite{MHV+09,BT09}.

In summary, we have demonstrated via x-ray absorption spectroscopy
the electronic origin of the magnetoelectric coupling in PZT/LSMO
multiferroic heterostructures, which arises from a change in the
valence state of Mn induced by electrostatic charge modulation. The
XANES data provide a direct and quantitative measure of the changes
in the number of electrons populating the Mn 3d $e_g$ bands, which
are responsible for the magnetic behavior of the LSMO. We conclude
that the large magnetoelectric coupling effect found in these
artificial heterostrutures results from an interfacial magnetic
reconstruction driven by charge accumulation. These findings show a
new pathway to the electronic control of spin in complex oxide
materials.

\begin{acknowledgments}
The authors acknowledge financial support by the NSF through MRSEC
DMR 0520495 (CRISP), FENA, and the NRI. Use of the Advanced Photon
Source was supported by the DOE, Office of Science, Office of Basic
Energy Sciences, under contract No. DE-AC02-06CH11357.
\end{acknowledgments}


\begin{thebibliography}{27}
\expandafter\ifx\csname
natexlab\endcsname\relax\def\natexlab#1{#1}\fi
\expandafter\ifx\csname bibnamefont\endcsname\relax
  \def\bibnamefont#1{#1}\fi
\expandafter\ifx\csname bibfnamefont\endcsname\relax
  \def\bibfnamefont#1{#1}\fi
\expandafter\ifx\csname citenamefont\endcsname\relax
  \def\citenamefont#1{#1}\fi
\expandafter\ifx\csname url\endcsname\relax
  \def\url#1{\texttt{#1}}\fi
\expandafter\ifx\csname urlprefix\endcsname\relax\def\urlprefix{URL
}\fi \providecommand{\bibinfo}[2]{#2}
\providecommand{\eprint}[2][]{\url{#2}}

\bibitem[{\citenamefont{Schmid}(1994)}]{Schmid94}
\bibinfo{author}{\bibfnamefont{H.}~\bibnamefont{Schmid}},
  \bibinfo{journal}{Ferroelectrics} \textbf{\bibinfo{volume}{162}},
  \bibinfo{pages}{317} (\bibinfo{year}{1994}).

\bibitem[{\citenamefont{Hill}(2000)}]{Hill00}
\bibinfo{author}{\bibfnamefont{N.~A.} \bibnamefont{Hill}}, \bibinfo{journal}{J.
  Phys. Chem. B} \textbf{\bibinfo{volume}{104}}, \bibinfo{pages}{6694}
  (\bibinfo{year}{2000}).

\bibitem[{\citenamefont{Ramesh and Spaldin}(2007)}]{RS07}
\bibinfo{author}{\bibfnamefont{R.}~\bibnamefont{Ramesh}} \bibnamefont{and}
  \bibinfo{author}{\bibfnamefont{N.~A.} \bibnamefont{Spaldin}},
  \bibinfo{journal}{Nature Mater.} \textbf{\bibinfo{volume}{6}},
  \bibinfo{pages}{21} (\bibinfo{year}{2007}).

\bibitem[{\citenamefont{Khomskii}(2009)}]{Khomskii09}
\bibinfo{author}{\bibfnamefont{D.~I.} \bibnamefont{Khomskii}},
  \bibinfo{journal}{Physics} \textbf{\bibinfo{volume}{2}}, \bibinfo{pages}{20}
  (\bibinfo{year}{2009}).

\bibitem[{\citenamefont{Eerenstein et~al.}(2006)\citenamefont{Eerenstein,
  Mathur, and Scott}}]{EMS06}
\bibinfo{author}{\bibfnamefont{W.}~\bibnamefont{Eerenstein}},
  \bibinfo{author}{\bibfnamefont{N.~D.} \bibnamefont{Mathur}},
  \bibnamefont{and} \bibinfo{author}{\bibfnamefont{J.~F.} \bibnamefont{Scott}},
  \bibinfo{journal}{Nature} \textbf{\bibinfo{volume}{442}},
  \bibinfo{pages}{759} (\bibinfo{year}{2006}).

\bibitem[{\citenamefont{Rondinelli et~al.}(2008)\citenamefont{Rondinelli,
  Stengel, and Spaldin}}]{RSS08}
\bibinfo{author}{\bibfnamefont{J.~M.} \bibnamefont{Rondinelli}},
  \bibinfo{author}{\bibfnamefont{M.}~\bibnamefont{Stengel}}, \bibnamefont{and}
  \bibinfo{author}{\bibfnamefont{N.~A.} \bibnamefont{Spaldin}},
  \bibinfo{journal}{Nature Nanotechnology} \textbf{\bibinfo{volume}{3}},
  \bibinfo{pages}{46} (\bibinfo{year}{2008}).

\bibitem[{\citenamefont{Fiebig}(2005)}]{Fiebig05}
\bibinfo{author}{\bibfnamefont{M.}~\bibnamefont{Fiebig}}, \bibinfo{journal}{J.
  Phys. D: Appl. Phys.} \textbf{\bibinfo{volume}{38}}, \bibinfo{pages}{R123}
  (\bibinfo{year}{2005}).

\bibitem[{\citenamefont{Thiele et~al.}(2007)\citenamefont{Thiele, D{\"o}rr,
  Bilani, R{\"o}del, and Schultz}}]{TDB+07}
\bibinfo{author}{\bibfnamefont{C.}~\bibnamefont{Thiele}},
  \bibinfo{author}{\bibfnamefont{K.}~\bibnamefont{D{\"o}rr}},
  \bibinfo{author}{\bibfnamefont{O.}~\bibnamefont{Bilani}},
  \bibinfo{author}{\bibfnamefont{J.}~\bibnamefont{R{\"o}del}},
  \bibnamefont{and} \bibinfo{author}{\bibfnamefont{L.}~\bibnamefont{Schultz}},
  \bibinfo{journal}{Phys. Rev. B} \textbf{\bibinfo{volume}{75}},
  \bibinfo{pages}{054408} (\bibinfo{year}{2007}).

\bibitem[{\citenamefont{Borisov et~al.}(2005)\citenamefont{Borisov, Hochstrat,
  Chen, Kleemann, and Binek}}]{BHC+05}
\bibinfo{author}{\bibfnamefont{P.}~\bibnamefont{Borisov}},
  \bibinfo{author}{\bibfnamefont{A.}~\bibnamefont{Hochstrat}},
  \bibinfo{author}{\bibfnamefont{X.}~\bibnamefont{Chen}},
  \bibinfo{author}{\bibfnamefont{W.}~\bibnamefont{Kleemann}}, \bibnamefont{and}
  \bibinfo{author}{\bibfnamefont{C.}~\bibnamefont{Binek}},
  \bibinfo{journal}{Phys. Rev. Lett.} \textbf{\bibinfo{volume}{94}},
  \bibinfo{pages}{117203} (\bibinfo{year}{2005}).

\bibitem[{\citenamefont{Laukhin et~al.}(2006)\citenamefont{Laukhin, Skumryev,
  Mart{\' \i}, Hrabovsky, S{\'a}nchez, Garc{\' \i}a-Cuenca, Ferrater, Varela,
  L{\"u}ders, Bobo et~al.}}]{LSM+06}
\bibinfo{author}{\bibfnamefont{V.}~\bibnamefont{Laukhin}},
  \bibinfo{author}{\bibfnamefont{V.}~\bibnamefont{Skumryev}},
  \bibinfo{author}{\bibfnamefont{X.}~\bibnamefont{Mart{\' \i}}},
  \bibinfo{author}{\bibfnamefont{D.}~\bibnamefont{Hrabovsky}},
  \bibinfo{author}{\bibfnamefont{F.}~\bibnamefont{S{\'a}nchez}},
  \bibinfo{author}{\bibfnamefont{M.~V.} \bibnamefont{Garc{\' \i}a-Cuenca}},
  \bibinfo{author}{\bibfnamefont{C.}~\bibnamefont{Ferrater}},
  \bibinfo{author}{\bibfnamefont{M.}~\bibnamefont{Varela}},
  \bibinfo{author}{\bibfnamefont{U.}~\bibnamefont{L{\"u}ders}},
  \bibinfo{author}{\bibfnamefont{J.~F.} \bibnamefont{Bobo}},
  \bibnamefont{et~al.}, \bibinfo{journal}{Phys. Rev. Lett.}
  \textbf{\bibinfo{volume}{97}}, \bibinfo{pages}{227201}
  (\bibinfo{year}{2006}).

\bibitem[{\citenamefont{Chu et~al.}(2008)\citenamefont{Chu, Martin, Holcomb,
  Gajek, Han, He, Balke, Yang, Lee, Hu et~al.}}]{CMH+08}
\bibinfo{author}{\bibfnamefont{Y.-H.} \bibnamefont{Chu}},
  \bibinfo{author}{\bibfnamefont{L.~W.} \bibnamefont{Martin}},
  \bibinfo{author}{\bibfnamefont{M.~B.} \bibnamefont{Holcomb}},
  \bibinfo{author}{\bibfnamefont{M.}~\bibnamefont{Gajek}},
  \bibinfo{author}{\bibfnamefont{S.-J.} \bibnamefont{Han}},
  \bibinfo{author}{\bibfnamefont{Q.}~\bibnamefont{He}},
  \bibinfo{author}{\bibfnamefont{N.}~\bibnamefont{Balke}},
  \bibinfo{author}{\bibfnamefont{C.-H.} \bibnamefont{Yang}},
  \bibinfo{author}{\bibfnamefont{D.}~\bibnamefont{Lee}},
  \bibinfo{author}{\bibfnamefont{W.}~\bibnamefont{Hu}}, \bibnamefont{et~al.},
  \bibinfo{journal}{Nature Mater.} \textbf{\bibinfo{volume}{7}},
  \bibinfo{pages}{478} (\bibinfo{year}{2008}).

\bibitem[{\citenamefont{Molegraaf et~al.}(2009)\citenamefont{Molegraaf,
  Hoffman, Vaz, Gariglio, van~der Marel, Ahn, and Triscone}}]{MHV+09}
\bibinfo{author}{\bibfnamefont{H.~J.~A.} \bibnamefont{Molegraaf}},
  \bibinfo{author}{\bibfnamefont{J.}~\bibnamefont{Hoffman}},
  \bibinfo{author}{\bibfnamefont{C.~A.~F.} \bibnamefont{Vaz}},
  \bibinfo{author}{\bibfnamefont{S.}~\bibnamefont{Gariglio}},
  \bibinfo{author}{\bibfnamefont{D.}~\bibnamefont{van~der Marel}},
  \bibinfo{author}{\bibfnamefont{C.~H.} \bibnamefont{Ahn}}, \bibnamefont{and}
  \bibinfo{author}{\bibfnamefont{J.-M.} \bibnamefont{Triscone}},
  \bibinfo{journal}{Adv. Mater.} \textbf{\bibinfo{volume}{21}},
  \bibinfo{pages}{3470} (\bibinfo{year}{2009}).

\bibitem[{\citenamefont{Burton and Tsymbal}(2009)}]{BT09}
\bibinfo{author}{\bibfnamefont{J.~D.} \bibnamefont{Burton}} \bibnamefont{and}
  \bibinfo{author}{\bibfnamefont{E.~Y.} \bibnamefont{Tsymbal}},
  \bibinfo{journal}{Phys. Rev. B} \textbf{\bibinfo{volume}{80}},
  \bibinfo{pages}{174406} (\bibinfo{year}{2009}).

\bibitem[{\citenamefont{Imada et~al.}(1998)\citenamefont{Imada, Fujimori, and
  Tokura}}]{IFT98}
\bibinfo{author}{\bibfnamefont{M.}~\bibnamefont{Imada}},
  \bibinfo{author}{\bibfnamefont{A.}~\bibnamefont{Fujimori}}, \bibnamefont{and}
  \bibinfo{author}{\bibfnamefont{Y.}~\bibnamefont{Tokura}},
  \bibinfo{journal}{Rev. Mod. Phys.} \textbf{\bibinfo{volume}{70}},
  \bibinfo{pages}{1039} (\bibinfo{year}{1998}).

\bibitem[{\citenamefont{Chmaissem et~al.}(2003)\citenamefont{Chmaissem,
  Dabrowski, Kolesnik, Mais, Jorgensen, and Short}}]{CDK+03a}
\bibinfo{author}{\bibfnamefont{O.}~\bibnamefont{Chmaissem}},
  \bibinfo{author}{\bibfnamefont{B.}~\bibnamefont{Dabrowski}},
  \bibinfo{author}{\bibfnamefont{S.}~\bibnamefont{Kolesnik}},
  \bibinfo{author}{\bibfnamefont{J.}~\bibnamefont{Mais}},
  \bibinfo{author}{\bibfnamefont{J.~D.} \bibnamefont{Jorgensen}},
  \bibnamefont{and} \bibinfo{author}{\bibfnamefont{S.}~\bibnamefont{Short}},
  \bibinfo{journal}{Phys. Rev. B} \textbf{\bibinfo{volume}{67}},
  \bibinfo{pages}{094431} (\bibinfo{year}{2003}).

\bibitem[{\citenamefont{Hong et~al.}(2003)\citenamefont{Hong, Posadas, Lin, and
  Ahn}}]{HPLA03}
\bibinfo{author}{\bibfnamefont{X.}~\bibnamefont{Hong}},
  \bibinfo{author}{\bibfnamefont{A.}~\bibnamefont{Posadas}},
  \bibinfo{author}{\bibfnamefont{A.}~\bibnamefont{Lin}}, \bibnamefont{and}
  \bibinfo{author}{\bibfnamefont{C.~H.} \bibnamefont{Ahn}},
  \bibinfo{journal}{Phys. Rev. B} \textbf{\bibinfo{volume}{68}},
  \bibinfo{pages}{134415} (\bibinfo{year}{2003}).

\bibitem[{\citenamefont{Hong et~al.}(2005)\citenamefont{Hong, Posadas, and
  Ahn}}]{HPA05}
\bibinfo{author}{\bibfnamefont{X.}~\bibnamefont{Hong}},
  \bibinfo{author}{\bibfnamefont{A.}~\bibnamefont{Posadas}}, \bibnamefont{and}
  \bibinfo{author}{\bibfnamefont{C.~H.} \bibnamefont{Ahn}},
  \bibinfo{journal}{Appl. Phys. Lett.} \textbf{\bibinfo{volume}{86}},
  \bibinfo{pages}{142501} (\bibinfo{year}{2005}).

\bibitem[{\citenamefont{Croft et~al.}(1997)\citenamefont{Croft, Sills,
  Greenblatt, Lee, Cheong, Ramanujachary, and Tran}}]{CSG+97}
\bibinfo{author}{\bibfnamefont{M.}~\bibnamefont{Croft}},
  \bibinfo{author}{\bibfnamefont{D.}~\bibnamefont{Sills}},
  \bibinfo{author}{\bibfnamefont{M.}~\bibnamefont{Greenblatt}},
  \bibinfo{author}{\bibfnamefont{C.}~\bibnamefont{Lee}},
  \bibinfo{author}{\bibfnamefont{S.-W.} \bibnamefont{Cheong}},
  \bibinfo{author}{\bibfnamefont{K.~V.} \bibnamefont{Ramanujachary}},
  \bibnamefont{and} \bibinfo{author}{\bibfnamefont{D.}~\bibnamefont{Tran}},
  \bibinfo{journal}{Phys. Rev. B} \textbf{\bibinfo{volume}{55}},
  \bibinfo{pages}{8726} (\bibinfo{year}{1997}).

\bibitem[{\citenamefont{Sub{\'\i}as et~al.}(1997)\citenamefont{Sub{\'\i}as,
  Garc{\'\i}a, Proietti, and Blasco}}]{SGPB97}
\bibinfo{author}{\bibfnamefont{G.}~\bibnamefont{Sub{\'\i}as}},
  \bibinfo{author}{\bibfnamefont{J.}~\bibnamefont{Garc{\'\i}a}},
  \bibinfo{author}{\bibfnamefont{M.~G.} \bibnamefont{Proietti}},
  \bibnamefont{and} \bibinfo{author}{\bibfnamefont{J.}~\bibnamefont{Blasco}},
  \bibinfo{journal}{Phys. Rev. B} \textbf{\bibinfo{volume}{56}},
  \bibinfo{pages}{8183} (\bibinfo{year}{1997}).

\bibitem[{\citenamefont{Bridges et~al.}(2001)\citenamefont{Bridges, Booth,
  Anderson, Kwei, Neumeier, Snyder, Mitchell, Gardner, and Brosha}}]{BBA+01}
\bibinfo{author}{\bibfnamefont{F.}~\bibnamefont{Bridges}},
  \bibinfo{author}{\bibfnamefont{C.~H.} \bibnamefont{Booth}},
  \bibinfo{author}{\bibfnamefont{M.}~\bibnamefont{Anderson}},
  \bibinfo{author}{\bibfnamefont{G.~H.} \bibnamefont{Kwei}},
  \bibinfo{author}{\bibfnamefont{J.~J.} \bibnamefont{Neumeier}},
  \bibinfo{author}{\bibfnamefont{J.}~\bibnamefont{Snyder}},
  \bibinfo{author}{\bibfnamefont{J.}~\bibnamefont{Mitchell}},
  \bibinfo{author}{\bibfnamefont{J.~S.} \bibnamefont{Gardner}},
  \bibnamefont{and} \bibinfo{author}{\bibfnamefont{E.}~\bibnamefont{Brosha}},
  \bibinfo{journal}{Phys. Rev. B} \textbf{\bibinfo{volume}{63}},
  \bibinfo{pages}{214405} (\bibinfo{year}{2001}).

\bibitem[{\citenamefont{Nagaev}(2001)}]{Nagaev01}
\bibinfo{author}{\bibfnamefont{E.~L.} \bibnamefont{Nagaev}},
  \bibinfo{journal}{Phys. Rep.} \textbf{\bibinfo{volume}{346}},
  \bibinfo{pages}{387} (\bibinfo{year}{2001}).

\bibitem[{\citenamefont{Lofland et~al.}(1997)\citenamefont{Lofland, Bhagat,
  Ghosh, Greene, Karabashev, Shulyatev, Arsenov, and Mukovskii}}]{LBG+97a}
\bibinfo{author}{\bibfnamefont{S.~E.} \bibnamefont{Lofland}},
  \bibinfo{author}{\bibfnamefont{S.~M.} \bibnamefont{Bhagat}},
  \bibinfo{author}{\bibfnamefont{K.}~\bibnamefont{Ghosh}},
  \bibinfo{author}{\bibfnamefont{R.~L.} \bibnamefont{Greene}},
  \bibinfo{author}{\bibfnamefont{S.~G.} \bibnamefont{Karabashev}},
  \bibinfo{author}{\bibfnamefont{D.~A.} \bibnamefont{Shulyatev}},
  \bibinfo{author}{\bibfnamefont{A.~A.} \bibnamefont{Arsenov}},
  \bibnamefont{and}
  \bibinfo{author}{\bibfnamefont{Y.}~\bibnamefont{Mukovskii}},
  \bibinfo{journal}{Phys. Rev. B} \textbf{\bibinfo{volume}{56}},
  \bibinfo{pages}{13705} (\bibinfo{year}{1997}).

\bibitem[{\citenamefont{Kirichok et~al.}(1985)\citenamefont{Kirichok, Kopaev,
  and Pashchenko}}]{KKP85}
\bibinfo{author}{\bibfnamefont{P.~P.} \bibnamefont{Kirichok}},
  \bibinfo{author}{\bibfnamefont{A.~V.} \bibnamefont{Kopaev}},
  \bibnamefont{and} \bibinfo{author}{\bibfnamefont{V.~P.}
  \bibnamefont{Pashchenko}}, \bibinfo{journal}{Russian Physics Journal}
  \textbf{\bibinfo{volume}{28}}, \bibinfo{pages}{983} (\bibinfo{year}{1985}).

\bibitem[{\citenamefont{Sikora et~al.}(2006)\citenamefont{Sikora, Kapusta,
  Kn\'{\i}\v{z}ek, Jir\'{a}k, Autret, Borowiec, Oates, Proch\'{a}zka, Rybicki,
  and Zajac}}]{SKK+06}
\bibinfo{author}{\bibfnamefont{M.}~\bibnamefont{Sikora}},
  \bibinfo{author}{\bibfnamefont{C.}~\bibnamefont{Kapusta}},
  \bibinfo{author}{\bibfnamefont{K.}~\bibnamefont{Kn\'{\i}\v{z}ek}},
  \bibinfo{author}{\bibfnamefont{Z.}~\bibnamefont{Jir\'{a}k}},
  \bibinfo{author}{\bibfnamefont{C.}~\bibnamefont{Autret}},
  \bibinfo{author}{\bibfnamefont{M.}~\bibnamefont{Borowiec}},
  \bibinfo{author}{\bibfnamefont{C.~J.} \bibnamefont{Oates}},
  \bibinfo{author}{\bibfnamefont{V.}~\bibnamefont{Proch\'{a}zka}},
  \bibinfo{author}{\bibfnamefont{D.}~\bibnamefont{Rybicki}}, \bibnamefont{and}
  \bibinfo{author}{\bibfnamefont{D.}~\bibnamefont{Zajac}},
  \bibinfo{journal}{Phys. Rev. B} \textbf{\bibinfo{volume}{73}},
  \bibinfo{pages}{094426} (\bibinfo{year}{2006}).

\bibitem[{\citenamefont{Shibata et~al.}(2003)\citenamefont{Shibata, Bunker, and
  Mitchell}}]{SBM03}
\bibinfo{author}{\bibfnamefont{T.}~\bibnamefont{Shibata}},
  \bibinfo{author}{\bibfnamefont{B.~A.} \bibnamefont{Bunker}},
  \bibnamefont{and} \bibinfo{author}{\bibfnamefont{J.~F.}
  \bibnamefont{Mitchell}}, \bibinfo{journal}{Phys. Rev. B}
  \textbf{\bibinfo{volume}{68}}, \bibinfo{pages}{024103}
  (\bibinfo{year}{2003}).

\bibitem[{\citenamefont{Bindu et~al.}(2005)\citenamefont{Bindu, Pandey, Kumar,
  Khalid, and Pimpale}}]{BPK+05}
\bibinfo{author}{\bibfnamefont{R.}~\bibnamefont{Bindu}},
  \bibinfo{author}{\bibfnamefont{S.~K.} \bibnamefont{Pandey}},
  \bibinfo{author}{\bibfnamefont{A.}~\bibnamefont{Kumar}},
  \bibinfo{author}{\bibfnamefont{S.}~\bibnamefont{Khalid}}, \bibnamefont{and}
  \bibinfo{author}{\bibfnamefont{A.~V.} \bibnamefont{Pimpale}},
  \bibinfo{journal}{J. Phys.: Condens. Matter} \textbf{\bibinfo{volume}{17}},
  \bibinfo{pages}{6396} (\bibinfo{year}{2005}).

\bibitem[{\citenamefont{Fang et~al.}(2000)\citenamefont{Fang, Solovyev, and
  Terakura}}]{FST00}
\bibinfo{author}{\bibfnamefont{Z.}~\bibnamefont{Fang}},
  \bibinfo{author}{\bibfnamefont{I.~V.} \bibnamefont{Solovyev}},
  \bibnamefont{and} \bibinfo{author}{\bibfnamefont{K.}~\bibnamefont{Terakura}},
  \bibinfo{journal}{Phys. Rev. Lett.} \textbf{\bibinfo{volume}{84}},
  \bibinfo{pages}{3169} (\bibinfo{year}{2000}).

\end{thebibliography}

\end{document}